**Title**: Martensitic-like transition between liquid crystalline and crystalline phases of prototypical discotic organic semiconductor

Nurjahan Khatun[a], Joe F. Khoury[b], Agnes C. Nkele[c], Lingyu Wang[a], Tieqiong Zhang[c], Partha P. Paul[d], Paul Chibuike Okoli[e], Nabila Shamim[e], Matteo Pasquali[b], and Kushal Bagchi[*a]

[a]Department of Chemistry, Rice University, Houston, Texas 77005, USA.

[b]Department of Chemical and Biomolecular Engineering, Rice University, Houston, Texas 77005, USA.

[c]Applied Physics Program-Smalley-Curl Institute, Rice University, Houston, Texas 77005, USA.

[d]Stanford Synchrotron Radiation Lightsource, SLAC National Accelerator Laboratory, Menlo Park, California 94025, USA.

[e]Department of Chemical Engineering, Prairie View A & M University, Prairie View, Texas 77446, USA.

[*]Email: kb122@rice.edu

**Significance:**

Martensitic transformations are ultrafast and reversible phase transitions that are leveraged across material science, from making steel harder to enabling shape memory alloys. The present understanding is that Martensitic transitions occur exclusively between crystalline solid phases. Challenging this view, we show here that an aligned columnar liquid crystal, a viscoelastic fluid, can transform to the crystalline state through an order preserving mechanism bearing close resemblance to a Martensitic transformation. Like the Austenite to Martensite transition in steel, a cooperative liquid crystal to crystal transition is only observed in rapid quench experiments. Our discovery of Martensitic-like transformations of liquid crystals could be leveraged to assemble macroscopically aligned organic semiconductor crystals for electronic applications.

**Abstract:**

Phase transitions between crystalline solids occur either through the nucleation and growth mechanism, a process that is slow and destructive or through the diffusion-less and order preserving Martensitic route. In both organic and inorganic materials, Martensitic transformations are known to occur only between phases with crystalline symmetry. We demonstrate here that for canonical discotic organic semiconductor HAT6, the transition between the liquid crystalline columnar hexagonal phase (Col$_H$) and the crystalline solid can occur through a mechanism that exhibits the hallmarks of Martensitic transformations: orientational correlations between parent and daughter phases, structural reversibility, and ultrafast kinetics. To access Martensitic-like solidification, the Col$_H$ phase of HAT6 is biaxially aligned in lithographically defined microchannels and crystallization is induced on deep supercooling. The transition mechanism is studied using a combination of polarized optical microscopy and X-ray scattering. At the largest accessible supercooling, the Col$_H$ - Crystal phase transition occurs at speeds of ~100 micrometer/s, a value that is seven orders of magnitude greater than the theoretical prediction for growth from isotropic melts. Our work suggests that Martensitic-like transformations can occur even between liquid crystals and crystals and are therefore more general than previously believed. Further, our work demonstrates that Martensitic-like transformations of anchored liquid crystals can be used to grow biaxially aligned crystals of organic molecules over arbitrarily long distances. As lattice alignment over large areas is desirable for devices like field-effect transistors and as several high-performance molecular semiconductors exhibit a $Col_H$ phase, our results hold general significance for organic electronics.

**Introduction**:

Understanding and controlling phase transition mechanisms are central themes of both condensed matter physics and materials science. The study of phase transitions can reveal striking similarities and analogies between material classes with vast apparent differences. A prominent example of phase transitions revealing hidden connections between materials systems is the liquid crystal-superconductor equivalence (1), studied initially by de Gennes. This analogy has guided the search for new liquid crystal (LC) phases: a pursuit for the parallel of Abrikosov phase in superconductors led to the discovery of the twist-grain boundary phase in LCs. Identifying similarities in phase behavior between hard and soft materials (2) is essential for arriving at a unified and predictive understanding of condensed matter.

Solid-solid transitions occur either through a 1) nucleation and growth or 2) Martensitic mechanism (3, 4). Nucleation and growth type processes are observed broadly across different physical states and are not specific to solid-state transformations. In contrast, the Martensitic mechanism is discussed exclusively in the context of solid-solid transitions with studies of these processes focusing mostly on metallic alloys. While conversions mediated by nucleation and growth tend to be irreversible, slow, and diffusion-limited, Martensitic phase changes are characterized by reversibility and rapid, diffusion-less kinetics. Martensite carbon steel can form from the Austenite phase at speeds of one kilometer per second (3). Further, while nucleation and growth transitions tend to be destructive, Martensitic transformations are displacive. When a parent single crystal undergoes a phase change through a nucleation and growth mechanism, the daughter phase is usually polycrystalline, and the initial order is permanently destroyed. Martensitic transformations which involve concerted atomic motions, can allow for reversible single crystal to single crystal phase changes. These properties of Martensitic processes have been exploited for applications, like shape memory elements (5).

Recent studies have shown that soft and organic materials, like metallic alloys, display Martensitic-like transformations. Several small-molecule organic semiconductor crystals, like 6,13-Bis(triisopropylsilylethynyl)pentacene (TIPS-pentacene), have shown to exhibit entirely solid-state disorder-order transitions that demonstrate the salient features of Martensitic transformations: orientational correlations between parent and daughter phases, reversibility, and fast kinetics (6-8). Martensitic-like transformations have also been reported for transitions between soft photonic crystals called blue phase liquid crystals (9-12). Blue phases either exhibit body-centered or simple cubic lattices, and recent studies have shown that single crystals of these phases can interconvert via a Martensitic-like mechanism. In both small-molecule organic crystals and blue phases, Martensitic transformations occur between phases with similar mechanical properties. In the disorder-order transitions in organic crystals, like textbook Martensitic transformations (13), both phases are elastic solids. In blue phases, the simple cubic to body-centered cubic transition occurs between materials that exhibit the mechanical response of liquids. It is unknown whether a fluid, like LC, can solidify through a cooperative, Martensitic-like mechanism. Such a transformation, in addition to expanding our fundamental understanding of cooperative phase transitions in condensed materials, would have broad technological implications. LCs can be easily assembled over macroscopic volumes, and a Martensitic-like transformation that would transfer molecular organization to three-dimensional crystals would result in entirely new ways to create macroscopically organized solids. Large volume molecular alignment in the solid state is vital for applications like organic electronics, where it leads to improved charge transport (14-25).

We show that the liquid crystalline columnar hexagonal ($Col_H$) phase of discotic mesogen HAT6 can transition to its crystal state through a cooperative and Martensitic-like mechanism during quench experiments. We demonstrate that when the $Col_H$ phase is aligned biaxially in microchannels and when the temperature is dropped rapidly, crystallization occurs with retention of order created in the precursor LC. At large supercooling, the $Col_H$ → Crystal transition occurs with a front velocity that is consistent with previous reports of Martensitic-like transformations in organic and soft materials (6, 10). Lastly, the transition is largely reversible. Our studies show that Martensitic-like transformations are not restricted to materials with similar mechanical properties but rather can also occur between viscoelastic liquids and solids.

## Methods

### Materials

HAT6 (2,3,6,7,10,11-Hexakis (hexyloxy) triphenylene) with an HPLC purity 98% was commercially procured from TCI (USA). 5CB (4-Cyano-4'-pentylbiphenyl) and 8CB (4-octyl-4'-cyanobiphenyl) liquid crystals were purchased from Sigma Aldrich (USA). Chloroform was procured from Sigma Aldrich (USA). A positive photoresist, AZ MIR 703 series, was procured from Merck (Germany). All chemicals were used as received.

### Fabrication of microchannel

Microchannels were fabricated using maskless photolithography (MLA 150). A positive photoresist (AZ MIR 703) was spin-coated onto a cleaned $SiO_2$ substrate at 3000 rpm for 60 seconds. Prior to coating, the substrates were sequentially cleaned in deionized water, acetone, and isopropanol (IPA) using ultrasonication. The cleaned substrates were subsequently subjected to oxygen plasma treatment for 10 minutes, followed by a 10-minute ozone treatment, to remove organic contaminants.

Photoresist-coated substrates were soft-baked at 90 °C for 1 minute to remove residual solvents. Photolithographic exposure was performed using a 375 nm light source at an exposure dose of 250 mJ/cm². A post-exposure bake was subsequently carried out at 110 °C for 2 minutes to stabilize the exposed pattern. The developed patterns were obtained using AZ MIF 300 developer, resulting in microchannels as shown in **Figure S1**, Supplementary Information (SI). The patterned region used for subsequent measurements measured 1 × 1 mm² for microscopy and 9 × 9 mm² for X-ray scattering.

*HAT6 spin-coating and thermal treatment*

Solutions of HAT6 in chloroform were prepared and deposited onto patterned substrates via spin-coating. Prior to deposition, the solution was vortexed for 10 minutes to ensure homogeneous mixing and then filtered through a 0.2 μm PTFE (polytetrafluoroethylene) membrane to remove particulates. The spin-coating was performed on ozone-treated microchannels at 3000 rpm for 30 seconds. After deposition, the films were annealed at 105 °C for 30 seconds to ensure complete melting of the sample into the isotropic phase. Following annealing, the films were cooled to 80 °C at a controlled rate of 10 °C/min to induce the Col$_H$ phase. Subsequent cooling to room temperature (25 °C) was carried out at various rates: quenching (by transferring the sample directly from a hot plate onto a copper block), 100 °C/min, 10 °C/min, 1 °C/min, and 0.1 °C/min, to induce the phase transition to the crystalline phase. Further, the cooling rate associated with quenching is estimated to be approximately 500 °C/min by fitting the anisotropy data across cooling rates ranging from 100 °C/min to 0.1 °C/min (see **Figure S2**). The reversibility of the phase transition was evaluated by reheating the crystalline state to the Col$_H$ phase at varying heating rates of 100 °C/min, 10 °C/min, 1 °C/min, and 0.1 °C/min.

*Surface topography and thickness measurements of the films*

The surface topography of bare and HAT6-filled microchannels (annealed) was characterized using atomic force microscopy (AFM) in tapping mode with a Park Systems NX20 instrument. Measurements were performed using a PPP-NCHR cantilever tip (force constant: 42 N/m; resonance frequency: 330 kHz). Trench depths and film thickness were determined from AFM line profiles as discussed in Section 1 and shown in **Figure S1** of the SI.

*Polarizing optical microscope (POM)*

*Phase sequence, transition temperatures, and in-plane anisotropy via quantitative image analysis:*

The phase sequence and phase transition temperatures of the films were identified using a POM (Olympus BX53M) equipped with a temperature-controlled THMS600-H Linkam hot stage. Textural images were captured *in situ* using a CCD camera (Olympus DP75) mounted on the microscope. Observations were performed by rotating the sample channels at 0° and 45° relative to the crossed polarizers to capture the textures. To correlate the orientations of the liquid crystalline and crystalline phases, the acquired POM images were quantitatively analyzed using ImageJ software. The in-plane anisotropy parameter is defined as:

$$In-plane\ anisotropy = \frac{I_{45} - I_0}{I_{45} + I_0} \qquad (1)$$

where $I_0$ and $I_{45}$ denote the image intensities measured at sample orientations of 0° and 45°, respectively. The difference between these intensities quantifies the degree of in-plane alignment, while the denominator serves as a normalization factor.

This analysis enabled a quantitative comparison of optical anisotropy between the liquid crystalline and crystalline states of the films.

*Isothermal growth kinetics at different supercooling temperatures:*

To determine the Col$_H$ to Crystal growth rate, time-resolved video recordings were acquired at different supercooling temperatures. The aligned Col$_H$ phase was cooled at a rate of 100 °C/min and then held isothermally at selected temperatures well below the melting temperature. Sequential frames were extracted at regular intervals (20 frames/s), and the position of the advancing growth front was tracked as a function of time. Crystal growth rate was obtained from the temporal progression of the growth front across multiple individual channels.

*Grazing incidence wide-angle X-ray scattering (GIWAXS)*

GIWAXS measurements were performed at beamline 11-3 of the Stanford Synchrotron Radiation Lightsource (SSRL), SLAC National Accelerator Laboratory, Stanford University. Diffraction patterns were collected using X-rays with a photon energy of 12.7 keV ($\lambda = 0.98$ Å). Measurements were conducted at an incidence angle of 0.14°, selected to be above the critical angle of the thin film but below that of the silicon substrate. Scattered X-rays were recorded on a Rayonix MX225 area detector (pixel size = 73.242 µm × 73.242 µm) positioned at a sample-to-detector distance of 309 mm. Each exposure was collected over 30 s with a total of five average scans.

*Rheology of HAT6*

To understand the viscoelastic properties of HAT6, rheological measurements were conducted using an advanced rheometric system (ARES G2, TA Instruments) equipped with a force-rebalanced transducer. Tests were performed in both steady-state and oscillatory modes. The powder sample was confined between two parallel plates in a standard parallel-plate geometry. The diameters of the top and bottom plates were 8 mm. The bottom plate was maintained at 100 °C, corresponding to the isotropic phase of HAT6. Once the sample was placed, the top plate was lowered to achieve a sample thickness of 0.9 mm. The system was then cooled at a rate of 5 °C/min to reach the $Col_H$ phase (70 °C) and subsequently the crystalline phase (40 °C) for rheological characterization. Strain and frequency sweeps were performed to determine the storage modulus ($G'$) and loss modulus ($G''$). Initially, a strain sweep was carried out at a fixed angular frequency of 5 rad/s to identify the linear viscoelastic regime (LVR) for the LC and crystalline states. Following this, a frequency sweep was conducted within the LVR, maintaining a constant strain amplitude of 0.01%. The frequency range for this measurement was 0.01–100 rad/s.

## Results:

*System of interest*

This study utilizes discotic liquid crystal (DLC), HAT6 (2,3,6,7,10,11-Hexakis (hexyloxy) triphenylene) (**Figure 1a**). HAT6 exhibits two well-defined phase transitions: one at ≈ 94 °C, corresponding to the transition from the $Col_H$ to the isotropic phase, and another at ≈ 67 °C, marking the transition from the crystalline phase to the $Col_H$ phase, as confirmed by DSC (Section S3, **Figure S3**, SI). In the $Col_H$ phase, HAT6 molecules self-assemble into columns, which in turn organize hexagonally in the $Col_H$. In the crystalline state, columnar structure is preserved, but hexagonal symmetry is lost, and cores are tilted with respect to the columnar axes. **Figure 1b** and **1c** present the molecular packing of HAT6 molecules in the $Col_H$ and crystalline states, respectively. We choose this system as previous X-ray scattering studies have suggested that when solidification occurs from the aligned $Col_H$ phase, the resulting crystalline solid exhibits biaxial anisotropy (26). The previous study, however, left several questions about the mechanism and kinetics of this order preserving phase transition unanswered.

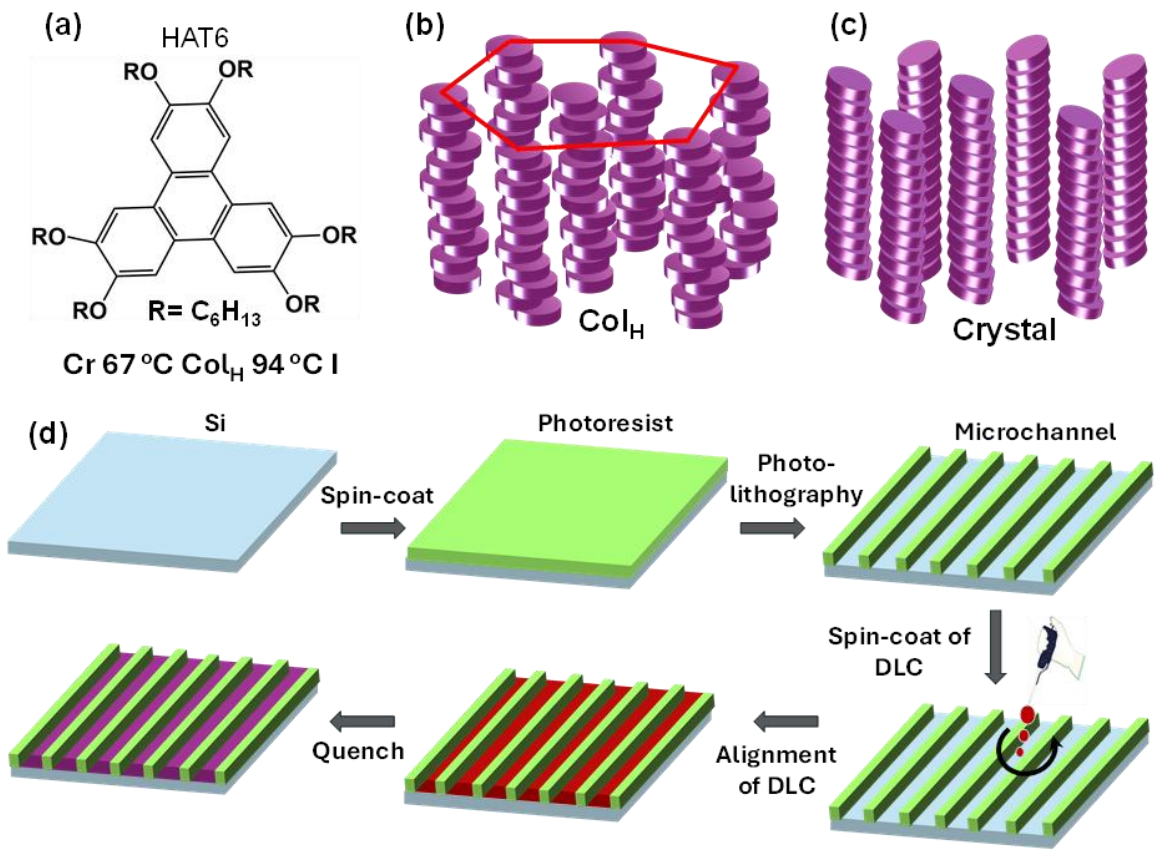

*Figure 1:* Chemical structure and phase behavior of HAT6 (a), molecular packing in liquid crystalline Col$_H$ (b) and crystalline phases (c), and workflow for microchannel-based alignment scheme (d). The width (w) and height (h) of the lithographically microfabricated channels in (d) are ~ 10 μm and ~ 1 μm, respectively.

*Alignment scheme of parent mesophase*

Microchannels fabricated via photolithography are employed to induce alignment of the Col$_H$ phase of discotic mesogen HAT6. Our lithographic workflow to prepare microchannels and confine HAT6 in them is shown in **Figure 1d**. Microchannels promote directional alignment with π-stacking perpendicular to the channel wall (27). POM images of the HAT6-filled microchannel are obtained during cooling from the isotropic phase. Images corresponding to the Col$_H$ phase, prepared through cooling at a rate of 10 °C/min, are shown in **Figure 2a**. The images are taken with the channels oriented at 0° and 45° relative to the crossed polarizers. The central square region corresponds to the patterned area, while the surrounding region is the un-patterned background. The square containing the microchannels exhibits complete extinction at 0° and maximum brightness at 45°, indicating directional in-plane alignment along the channel axis. The organization of the Col$_H$ phase in the microchannels is schematically presented in **Figure 2d**.

Microchannels induce biaxial alignment owing to the interfacial orientation preferences of discotic molecules. Discotic molecules prefer to be oriented with a face-on orientation with respect to a solid interface, while at the air interface, the molecules tend to adopt edge-on orientation with the discotic cores parallel to the surface normal (28-30). In open films, competition arises between these two interfacial anchoring tendencies. When the film is sufficiently thin edge-on packing dominates, driven by the stronger influence of the air-LC interface; in discotic systems, surface tension tends to be significantly greater than interfacial tension. In open films, while there is preferential edge-on packing, the columnar axis exhibits isotropic in-plane organization. Microchannel confinement, through creating two new vertical surfaces, creates a favored in-plane direction (26, 27). The discs stack perpendicular to the wall, resulting in an energetically favorable configuration at the free surface and the two side walls. Creating an aligned precursor state allows us to study the $Col_H$ → Crystal phase transition in the absence of defects.

### *Orientational correlations between parent and daughter phases in microchannels*

Dropping the temperature rapidly after alignment of the $Col_H$ phase results in a phase transition that preserves biaxial alignment. The image shown in **Figure 2b** is captured after the $Col_H$ phase is quenched to room temperature, resulting in a phase transition to the crystalline phase. A similar in-plane molecular orientation is observed in the crystalline state, suggesting that the alignment is preserved during the phase transition from $Col_H$ to the crystalline state. This behavior is consistently reproduced across three independent samples, demonstrating robust orientational correlation between the liquid crystalline parent phase and the resulting crystalline daughter phase (**Figure S5**). Our results are consistent with previous X-ray studies reporting preservation of structural order during the $Col_H$-to-crystal transition of HAT6 confined in microchannels. However, the present work extends beyond earlier reports by providing quantitative correlations between the two phases (**Figure 2c, d**). The results clearly demonstrate that optical anisotropy is maximized in both phases when the channel is oriented at ±45° relative to the crossed polarizers and gradually decreases to a minimum as the orientation approaches 0° (**Figure 2c**). The strong degree of orientation correlation between the parent liquid crystalline $Col_H$ phase and the daughter crystalline phase is consistent with a martensitic-like transformation mechanism. The direction of the columns in the liquid crystalline and crystalline states is schematically represented in **Figure 2e, f.**

The retention of liquid crystalline optical texture in the crystalline state appears to be independent of pattern geometry. When the $Col_H$ phase is confined to concentric circular microchannels, alternating bright and dark regions arise due to continuous rotation of the director in the plane (see **Figure S6)**. Interestingly, when the $Col_H$ phase is quenched and a crystal is formed, this "Maltese-cross" texture is retained. The alternating bright and dark regions in the crystal are produced through continuous rotation of the crystalline lattice within the walls of the circular microchannels.

### *Selection of order preserving $Col_H$ to crystal transitions*

Ultrafast kinetics is one of the salient features of Martensitic transformations (31, 32) . If the order preserving and displacive transition shown in **Figure 2a** is a Martensitic-like transformation, it should only be observed at the fastest cooling rates. To test the rigor of our analogy, we induce crystallization under a range of cooling conditions, including 500 °C/min, 100 °C/min, 10 °C/min, 1 °C/min, and 0.1 °C/min. The corresponding POM images are shown in **Figure 3**. The images reveal that the crystalline phase retains directional planar alignment when cooled rapidly, while this alignment diminishes progressively at slower cooling rates. Notably, at the slowest cooling rate of 0.1 °C/min, anisotropy is highest when measured at 0° and lowest at 45°, indicating a distinct reorganization of molecular orientation under these conditions. The crystalline organization observed at the slowest rate is the thermodynamically favored structure and involves restructuring the molecular organization in the parent $Col_H$ phase. The fast-cooling rate, which retains the columnar orientation of the LC phase, is kinetically favored and trapped through quenching.

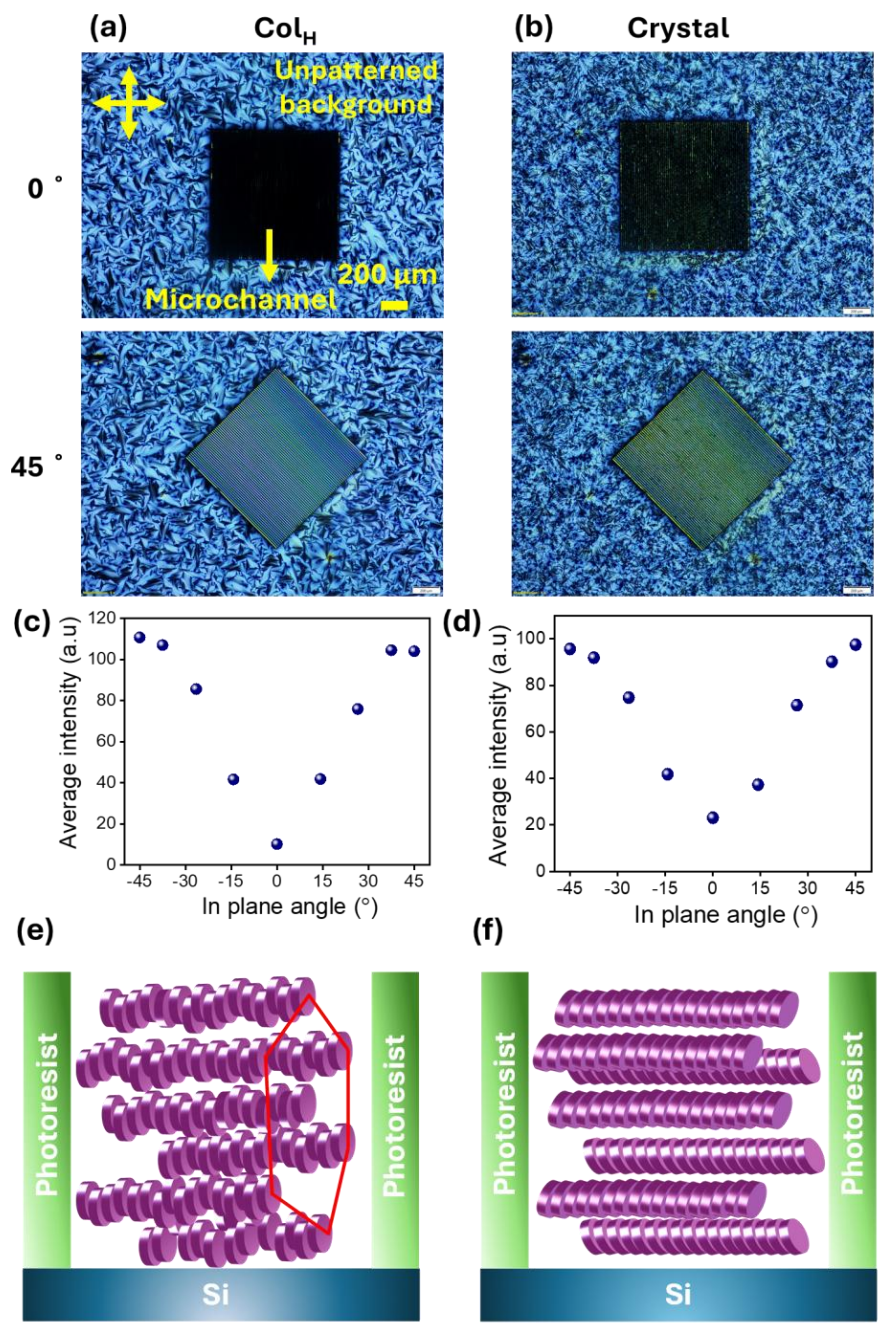

*Figure 2:* Growth of aligned crystals of HAT6 through Martensitic-like route. Polarized optical microscopy images of HAT6 in its Col$_H$ (a) and crystal states (b). Images are acquired with channels oriented along and rotated 45° with respect to cross-polarizers. The central square region corresponds to the region with microchannels, while the surrounding region is the unpatterned background. Variation of intensity in microchannels with in-plane rotation in Col$_H$ (c) and crystal states (d) of HAT6. Schematic representation of molecular packing in microchannels in Col$_H$ (e) and crystal states (f). There are clear orientational correlations between the parent Col$_H$ and daughter crystal phases.

Change in intensity with in-plane rotation can be used to quantitatively determine in plane anisotropy. To quantify biaxial optical anisotropy, POM images are integrated within the patterned region at 0° and 45°, which is described in detail in the Method section. The difference in intensity between these two positions determines the degree of in-plane anisotropy. The sum of the intensities at the two angles ($I_{45}$ + $I_0$) is used as a normalization factor. The metric ($I_{45}$ - $I_0$)/($I_{45}$ + $I_0$) equals 1 for perfect biaxial alignment and 0 when molecules are completely random in the plane. The resulting anisotropy values for various cooling rates are plotted in **Figure 3f**. In the $Col_H$ phase, the anisotropy reaches 90%, reflecting strong directional order. Upon transition to the crystalline phase, the anisotropy remains high (~70%) during quenching but decreases steadily with slower cooling rates, demonstrating the critical role of cooling rate in retaining the organization of the parent phase. These results suggest that under quenching, the $Col_H$-to-crystal transition proceeds via a Martensitic-like pathway, while slower cooling results in a more destructive transition akin to the more common nucleation-and-growth mechanism.

### *Reversibility of order preserving $Col_H$ to crystal transition*

Martensitic transformations tend to be reversible (33-35). To assess reversibility, the crystalline phase is reheated to the $Col_H$ phase at a heating rate of 100 °C/min. POM images are captured at three stages: the $Col_H$ phase at 80 °C (**Figure 3a**), the crystalline phase at 25 °C (**Figure 3d**), and the re-formed $Col_H$ phase at 80 °C during reheating (**Figure 3e**). The images clearly demonstrate that the alignment in the $Col_H$ phase is largely preserved through the transition to and from the crystalline phase, indicating reversibility. To further quantify this behavior, the POM images were integrated for samples subjected to reheating at different rates: 100 °C/min, 10 °C/min, and 1 °C/min. The analysis reveals that the degree of reversibility is largely independent of the heating rate (**Figure S7**). A reduction in anisotropy during the second heating cycle (marked by the red dotted line in **Figure 3f**) is likely due to tempering. Quenching locks in a non-equilibrium crystalline organization and raising the temperature provides the system with energy to evolve towards a more stable configuration before the transformation into the liquid crystalline phase.

### *Isothermal transformation kinetics*

To compare the transformation kinetics observed here with models for molecular crystal growth rate from supercooled liquids, we perform isothermal measurements. Here, we rapidly drop the temperature from 80 °C, where the system is in the $Col_H$ phase, to different temperatures below the melting point. The front velocity is calculated at different supercooling temperatures ($T_s$). Sequential frames are extracted at regular time intervals, and the position of the growth front is tracked as a function of time (**Figure S8**). The front velocity is determined from the total front displacement over the growth time and averaged across four independent channels. Front velocity is plotted against the degree of supercooling, which can be defined as $T_{Crystal-ColH} - T_s$, and is presented in **Figure 4a**. The front velocity increases gradually with increasing supercooling as the temperature moves further away from the melting temperature (Crystal → Col$_H$). At temperatures well below the Col$_H$ → Crystal transition, the velocity reaches $10^{-4}$ m/s. These observations clearly indicate that the phase transition is inherently rapid.

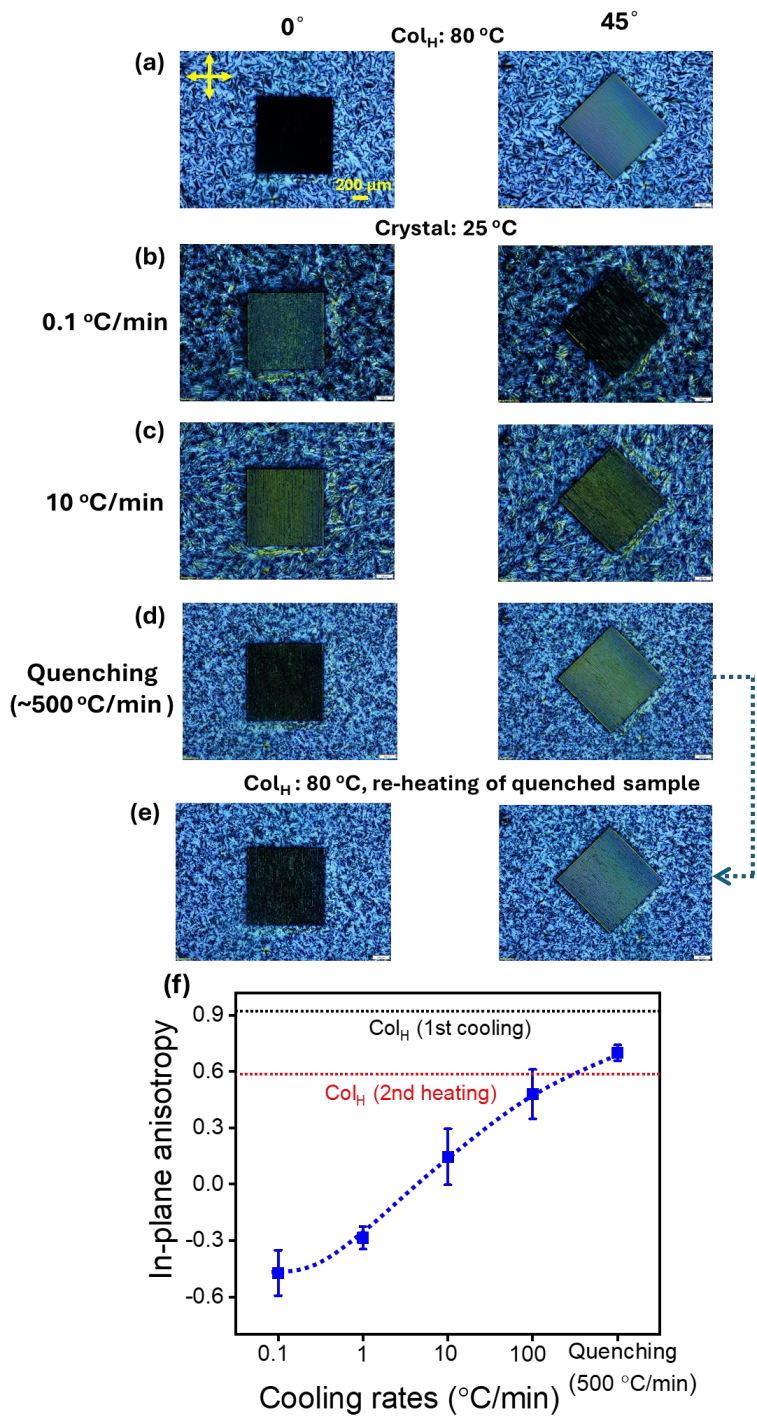

*Figure 3:* *Cooling rate dependence of anisotropy transfer from Col$_H$ phase to crystal state. POM images of the aligned Col$_H$ (a), crystalline phase obtained at various cooling rates (b-d) and re-heating the quench sample to Col$_H$ phase (e). In-plane anisotropy, quantified as the intensity difference between 0° and 45° orientations within the patterned square region and in the crystal state as a function of cooling rates (f). Values are averaged over three independent samples; error bars represent the standard deviation. Quenching the sample leads to order preserving phase transitions, while slower cooling rates lead to reconstructive transformations.*

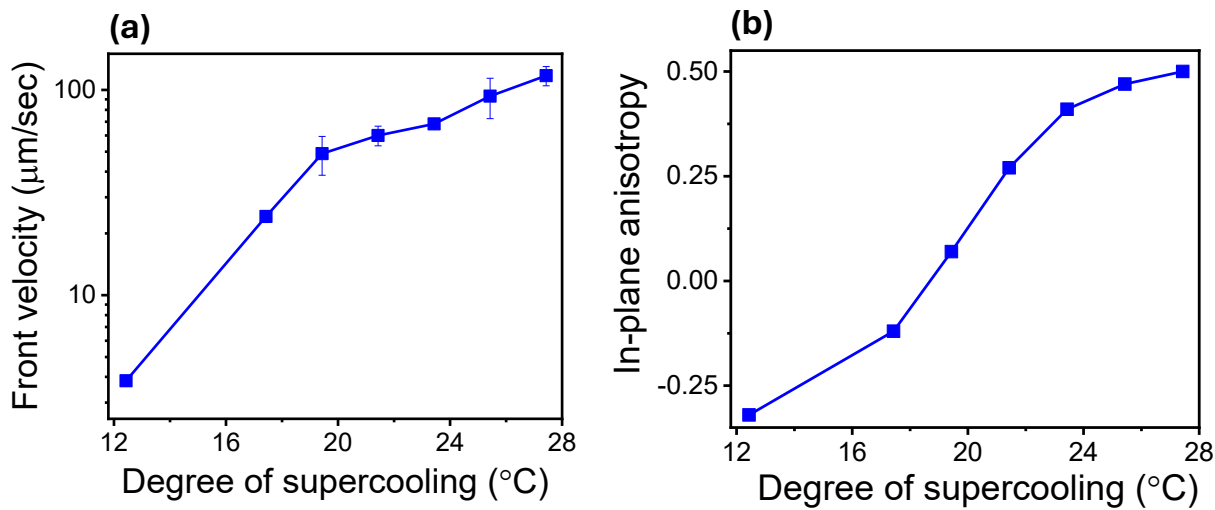

***Figure 4:*** *Kinetics of the Col$_H$ → Crystal transition in microchannels. Isothermal crystal growth front velocity (a) and in-plane anisotropy (b), as a function of the degree of supercooling. Error bars in (a) represent the standard deviation across the four channels. Deep supercooling of the Col$_H$ phase results in phase transitions that are both fast and order preserving.*

We first compare the observed transformation speeds against the Burke-Broughton-Gilmer (BBG) model, a widely used theoretical framework for predicting crystal growth velocities from supercooled molecular liquids (36). The BBG model predicts that the growth front speed, G(T), is:

$$G(T) = \frac{D(T)}{\alpha}[1 - e^{\frac{-\Delta G(T)}{K_B T}}]e^{\frac{-\Delta S_m}{R}} \quad (2)$$

Where D is the diffusion coefficient, α is the molecular length, $\Delta S_m$ is the entropy of fusion, R is the universal gas constant, and ΔG is the free-energy difference, which is calculated using the expression:

$$\Delta G(T) = \frac{\Delta H \Delta T}{T_m} \quad (3)$$

Where ΔH is the enthalpy of fusion, $T_m$ is the melting temperature, and ΔT is the difference between $T_m$ and crystallization temperatures ($T_s$) (see **Figure S9**). The parameter D is estimated to be $10^{-14}$ m$^2$/s based on values reported for HAT5 in the Col$_H$ phase (37). For HAT6, α is taken as approximately 2.0 nm (38). The parameters ΔH, ΔS$_m$, and T$_m$ are obtained from DSC data acquired during heating at a rate of 10 °C/min (**Figure S3**). The resulting values are ΔH = 37.33 kJ/mol, ΔS$_m$= 109.6 J/mol/K and T$_m$ = 67.5 °C. The crystallization temperature, T$_s$, is held at 40 °C. Using these parameters, the BBG model predicts a crystal growth rate on the order of $10^{-11}$ m/s, which is seven orders of magnitude lower than the experimentally observed value of $10^{-4}$ m/s at the largest supercooling in **Figure 4a**. Even at low supercooling, the model is off by several orders of magnitude. While it is unsurprising that a theory developed to model crystal growth from isotropic liquids fails for LCs, the degree of deviation indicates a completely different approach is required to model the kinetics of the Col$_H$ → Crystal transition.

The experimental growth rate at the largest supercooling is compared with another theoretical framework, the Wilson-Frenkel (WF) model (39) (Equation 4):

$$G(T) = \frac{D(T)}{\alpha}\left[1 - e^{\frac{-\Delta G(T)}{K_B T}}\right] \quad (4)$$

With the same parameters, the Wilson-Frenkel model predicts a growth velocity on the order of $10^{-6}$ m/s, still two orders of magnitude lower than the experimental value. Traditional models of crystal growth from melts cannot account for the crystallization rates observed at large supercooling. A new theoretical framework is required to model crystallization speeds from structured and anisotropic fluids.

Martensitic transformations exhibit diffusion-less kinetics. For the Col$_H$ → Crystal transformation to be diffusion-less, the growth front must move faster than molecules can diffuse. The direction in which we observe the growth front is perpendicular to the direction of π-stacking. In this direction, the spacing between molecules is approximately one molecular diameter, which ~ 2 nm. The fundamental unit a crystal most grow by in the direction of observation is therefore ~ 2 nm. We estimate the time (t) molecules would take to diffuse one diameter, $\alpha$, using the equation:

$$t = \frac{\alpha^2}{2D} \quad (5)$$

Setting the diffusion coefficient D = $10^{-14}$ m²/s (37), we calculate that it takes 200 μs for molecules in the Col$_H$ phase to diffuse a distance of one molecular diameter. Based on the observed growth front of $10^{-4}$ m/s at the largest supercooling (**Figure 4**), we estimate it takes 20 μs for the interface to grow one molecular diameter. Our observed transformation speed is therefore consistent with a diffusion-less transition. We note that the value we use for the diffusion coefficient is measured at a temperature above the Col$_H$ → Crystal transition. We are observing the transition at considerable supercooling. The value of D = $10^{-14}$ m²/s is therefore a conservative upper bound estimate. The diffusion-coefficient of HAT6 at the largest supercooling is likely to be much smaller, and the growth front is therefore faster than diffusion by at-least an order of magnitude.

The speed of the phase boundary during the LC to crystal transition and the mechanism of transformation are closely correlated. **Figure 4b** shows that deep supercooling not only induces a fast transition but also selects a mechanism that leads to preservation of precursor phase order. At the largest supercooling, the transition mechanism is clearly Martensitic-like. In contrast, small supercooling induces slow and reconstructive transitions that produce crystals that, unlike Col$_H$ exhibit extinction under cross-polarizers at 45° rather than 0° (**Figure S10**). As expected, large and small degrees of supercooling during isothermal transformations have the same effect on crystalline structure as cooling at fast and slow rates, respectively. Like cooling at 0.1 °C/min, the lowest degree of supercooling in **Figure 4b** generates crystals that exhibit negative in-plane anisotropy values. Analogous to rapid quenching, crystals grown isothermally at the lowest accessible temperatures produce crystals that exhibit anisotropy values closest to the aligned $Col_H$ phase.

*Molecular packing in crystalline films formed from Martensitic mechanism*

We performed grazing incidence wide angle scattering (GIWAXS) to probe molecular packing (40, 41) in crystalline films prepared from cooling the aligned $Col_H$ phase. Shown in **Figure 5** are GIWAXS patterns collected from biaxially aligned HAT6 crystals prepared by quenching; X-Ray patterns from films prepared at slower cooling rates are shown in **Figure S11**. $Q_z$ is the scattering vector in the out-of-plane direction while $Q_y$ is orthogonal both to the surface normal and the X-ray beam. GIWAXS patterns are collected with the X-Ray parallel (a) and perpendicular (b) to the channels. When the X-Ray is parallel a series of rings are observed in the plane between 1.4 and 1.9 Å$^{-1}$. This is the range in which the π-stacking peaks are expected to appear. These peaks disappear when the sample is rotated 90°, establishing that the direction of π-stacking is perpendicular to the channel walls in the crystalline state. Integration of the two-dimensional patterns over the Q range of 0-1.8 Å$^{-1}$ and χ (where χ is the azimuthal angle, with χ = 0° is along the $Q_z$ direction and χ = 90° is along the $Q_Y$ direction) range of 70°-80° for both beam orientations further corroborates this conclusion, showing pronounced π-stacking peaks only when the X-ray beam is parallel to the channels (**Figure S12**). Further, the scattering patterns confirm that the films prepared from quenching the $Col_H$ are indeed crystalline. Quenching could also induce the formation of a liquid crystalline glass; the sharp scattering peaks in **Figure 5** unambiguously establish that crystallization occurs even at the fastest cooling rate employed in this study.

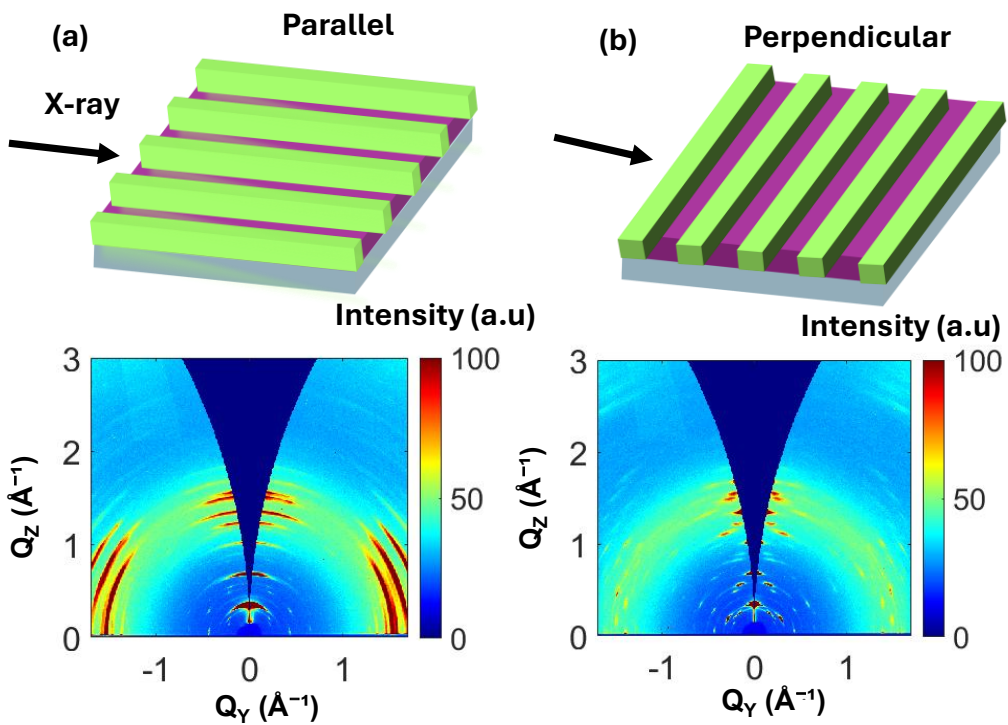

*Figure 5:* *X-ray characterization of quenched HAT6 crystal in microchannels. Patterns are collected with the microchannels oriented parallel (0°) (a) and perpendicular (90°) (b) to the incident X-ray beam. The colors in the GIWAXS image represent scattered intensity, with red and blue corresponding to high and low counts, respectively. Variations in scattering intensity with in-plane rotation confirm biaxial alignment in the crystalline state, with the direction of π-π stacking perpendicular to the walls.*

## Viscoelasticity of Col$_H$ and crystal phases

Textbook Martensites and their precursor phases are Hookean solids. We observe here that the Col$_H$ to crystal transition in HAT6 proceeds through a Martensitic-like mechanism, suggesting that these transformations are not restricted to elastic solids. While HAT6 in its Col$_H$ phase is not a Hookean solid, neither is it a Newtonian liquid. Characterizing the viscoelastic behavior of both the Col$_H$ and crystalline phases is necessary to determine the mechanical and rheological attributes of materials that can exhibit Martensitic-like transformations. To this end, we studied the rheology of the Col$_H$ and crystal states of HAT6. We performed shear rheometry on a bulk powder sample (thickness ≈ 0.9 mm), as thin film ((~800 nm) measurements are infeasible. Although these measurements were conducted on relatively thick samples, the viscoelastic characteristics of the LC phase are expected to be similar in thin films. The frequency and strain sweep in both crystalline and liquid crystalline states are shown in **Figure 6** and **Figure S13**, respectively.

The liquid crystalline Col$_H$ phase of HAT6 exhibits the rheological response of a textbook viscoelastic liquid. The loss modulus (G″) exceeds the storage modulus (G′) at low frequencies indicating dominant viscous behavior at long timescales. At high frequencies G′ surpasses G″ indicating that at short time-scales elastic behavior becomes more prominent (**Figure 6a**). The crystalline state of HAT6 displays the rheological behavior of a viscoelastic solid, with G′ higher than G″ across the frequency range (**Figure 6b**). Unlike a purely elastic solid, however, the value of G″ is appreciable and is within one order of magnitude of G′ at almost all frequencies. Both the Col$_H$ and crystalline phase exhibit substantial viscoelasticity. The combination of structural data reported in **Figures 2-5** and the rheology data in **Figure 6** establishes that Martensitic transformations can occur between structured viscoelastic liquids and viscoelastic crystalline solids.

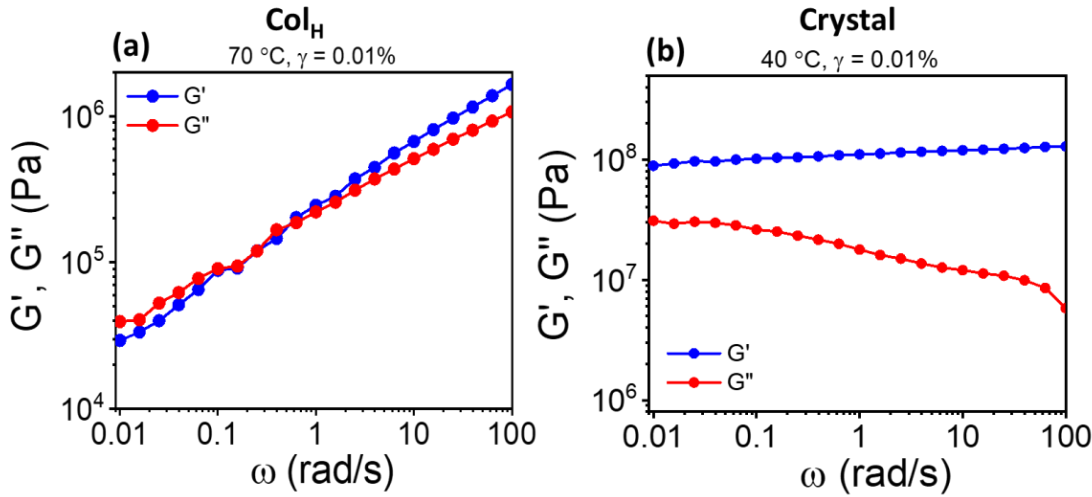

***Figure 6:** Shear rheometry of HAT6 in Col$_H$ and crystal states. Storage (G') and loss (G") modulus in Col$_H$ phase (a) and crystal states (b). For both measurements, 0.01 strain percentage is used. Col$_H$ and crystal states are accessed by performing measurements at 70 and 40 °C, respectively. The Col$_H$ and crystal states exhibit the rheological signatures of a viscoelastic liquid and a viscoelastic solid, respectively.*

**Discussion**

The work presented here is the first example, to the best of our knowledge, of a cooperative, Martensitic-like transformation occurring in a phase transition involving loss of a translational degree of freedom. The Col$_H$ phase is crystalline in two-dimensions but has translational freedom in the axis perpendicular to the hexagonal lattice. Upon cooling of the Col$_H$ phase of HAT6 and forming a solid with full three-dimensional periodicity, all translational degrees of freedom are lost. Our work establishes that Martensitic-like processes are not restricted to transformations between phases with three-dimensional periodicity. Preliminary results suggest that displacive, order-retaining phase transition mechanisms may be inaccessible for transformations involving loss of multiple translational degrees of freedom. Shown in Section 13 of the SI are results on crystallization from nematic (**Figure S14**) and smectic (**Figure S15**) precursor phases. For comparison with the results shown here, we confine nematic and smectic formers 5CB and 8CB in a rubbed LC cell that promotes directional planar anchoring. The aligned LC phase is cooled at 100 °C/min to induce rapid crystallization. Upon solidification from nematic and smectic phases, the resulting crystal shows no retention of the parent liquid-crystalline order. While the precursor nematic and smectic phases show biaxial molecular alignment as evidenced by intensity changes on rotating the sample in-the plane under a POM microscope, the crystalline sample is equally bright at all rotation angles. These observations indicate that cooperative, Martensitic-like transitions might be restricted to phase transformations involving the loss of no more than one degree of translational freedom. Another possibility is that transfer of liquid crystalline order to the solid state from the nematic and smectic phases requires significantly higher cooling rates than employed for the Col$_H$ phase. Future studies that induce crystallization of nematic and smectic LCs under even faster rates or under even deeper supercooling can test this hypothesis.

Cooperative transitions between the Col$_H$ and crystal phases are possible only when other competing solidification mechanisms can be suppressed. Previous studies have reported that upon cooling, the Col$_H$ phase can either form a liquid crystalline glass (24, 42-44) or crystallize through a nucleation and growth mechanism (45, 46). The three possible solidification mechanisms are shown in **Figure 7**. When cooled sufficiently fast all LCs are expected to vitrify. Likewise, our study establishes that slower cooling rates result in transformations consistent with a nucleation and growth mechanism. The Martensitic route is therefore possible when the cooling rate is fast enough to avoid nucleation and growth but not so fast that crystallization is averted altogether. Our study establishes that this window can be accessed by cooling the Col$_H$ phase of HAT6 at rates of ~100 °C/min. Future studies can establish whether similar rates can be employed for other mesogens that exhibit a Col$_H$ phase.

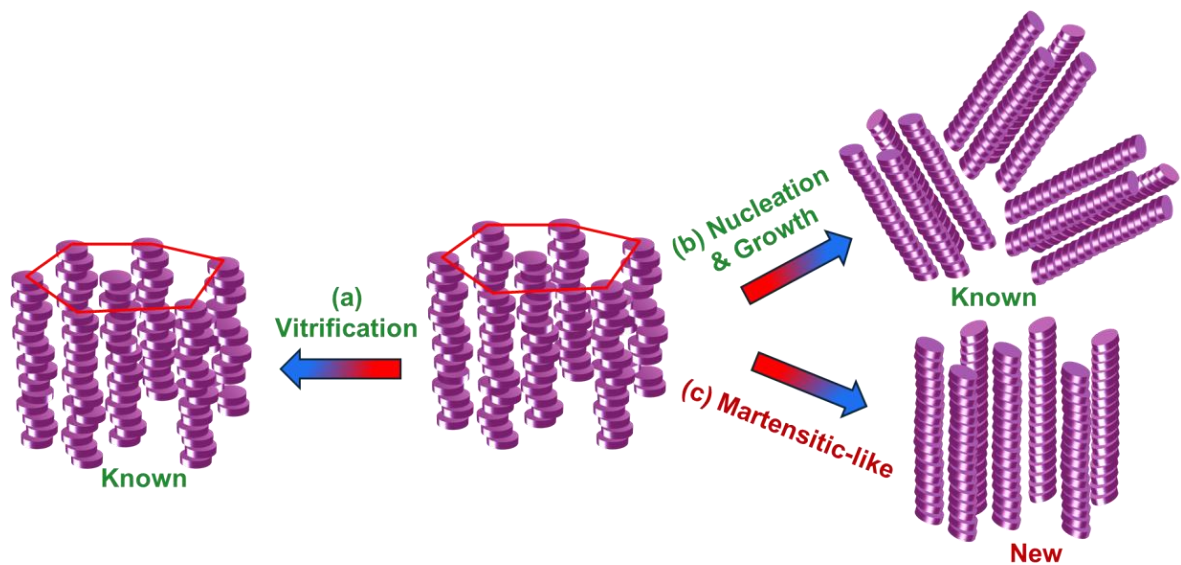

***Figure 7:*** *Molecular packing when solidification of the Col$_H$ state occurs via vitrification (a), crystal nucleation and growth (b), and Martensitic-like mechanisms (c). The arrow represents cooling. While vitrification (a) and crystallization through nucleation and growth (b) have been reported previously, the Martensitic-like mechanism (c) has been reported here for the first time for the Col$_H$ → Crystal phase transition.*

Cooperative crystallization of the aligned Col$_H$ phase holds promise for organic electronics applications. Owing to its fluidity, the Col$_H$ phase can be easily aligned over macroscopic dimensions (14, 29, 47, 48), and our work demonstrates that under appropriate processing conditions, its anisotropy can be transferred to the crystalline state. Several well-studied small-molecule organic semiconductors, like phthalocyanines (27), and hexabenzocoronenes (49, 50), are crystals at room temperatures but exhibit a Col$_H$ phase at elevated temperatures. Alignment of the precursor LC phase and transfer of order could therefore be used to grow organic semiconductor crystals over large volumes for organic electronic devices like organic field effect transistors, which typically operate at room temperature. Large area crystalline alignment is key for optimizing charge carrier mobility and thereby enhancing device performance (25, 49, 51-56).

In summary, we show that on sufficiently large supercooling the transition between the Col$_H$ and crystalline states of organic semiconductor HAT6 can proceed through a mechanism that preserves the biaxial anisotropy of the precursor liquid crystalline phase. This new mechanism bears close resemblance to a Martensitic transition, which is usually discussed in the context of solid-solid transitions. Cooperative, Martensitic-like transitions might therefore be more general than suggested in material science textbooks (13). Further we observe that irrespective of the degree of supercooling and the transformation mechanism, the Col$_H$ → Crystal transition occurs at speeds that cannot be accounted for by existing theories of crystal growth. Our work presents a challenge to theorists to generate models of crystal growth that incorporate the new physics that emerges when crystallization occurs from a liquid crystal rather than an isotropic liquid.

# Supplementary Information

**Title**: Martensitic-like transition between liquid crystalline and crystalline phases of prototypical discotic organic semiconductor

Nurjahan Khatun[a], Joe F. Khoury[b], Agnes C. Nkele[c], Lingyu Wang[a], Tieqiong Zhang[c], Partha P. Paul[d], Paul Chibuike Okoli[e], Nabila Shamim[e], Matteo Pasquali[b], and Kushal Bagchi[*a]

[a]Department of Chemistry, Rice University, Houston, Texas 77005, USA.

[b]Department of Chemical and Biomolecular Engineering, Rice University, Houston, Texas 77005, USA.

[c]Applied Physics Program-Smalley-Curl Institute, Rice University, Houston, Texas 77005, USA.

[d]Stanford Synchrotron Radiation Lightsource, SLAC National Accelerator Laboratory, Menlo Park, California 94025, USA.

[e]Department of Chemical Engineering, Prairie View A & M University, Prairie View, Texas 77446, USA.

[*]Email: kb122@rice.edu

## Section S1: AFM profiles and line cuts to determine surface topography and thickness of sample

Show in **Figure S1** are AFM images and corresponding linecut profiles of the bare microchannel (a,b) and the HAT6-filled microchannel after annealing (c,d). The reduction in trench height observed in the line profile of the HAT6-filled film (d) confirms that material from the top of the channel region flows into the trench region during thermal treatment. From the line-cut profiles, the trench heights of the bare and annealed HAT6-filled microchannels are approximately 1500 nm and 630 nm, respectively, yielding a film thickness of about 870 nm (the difference between the two heights).

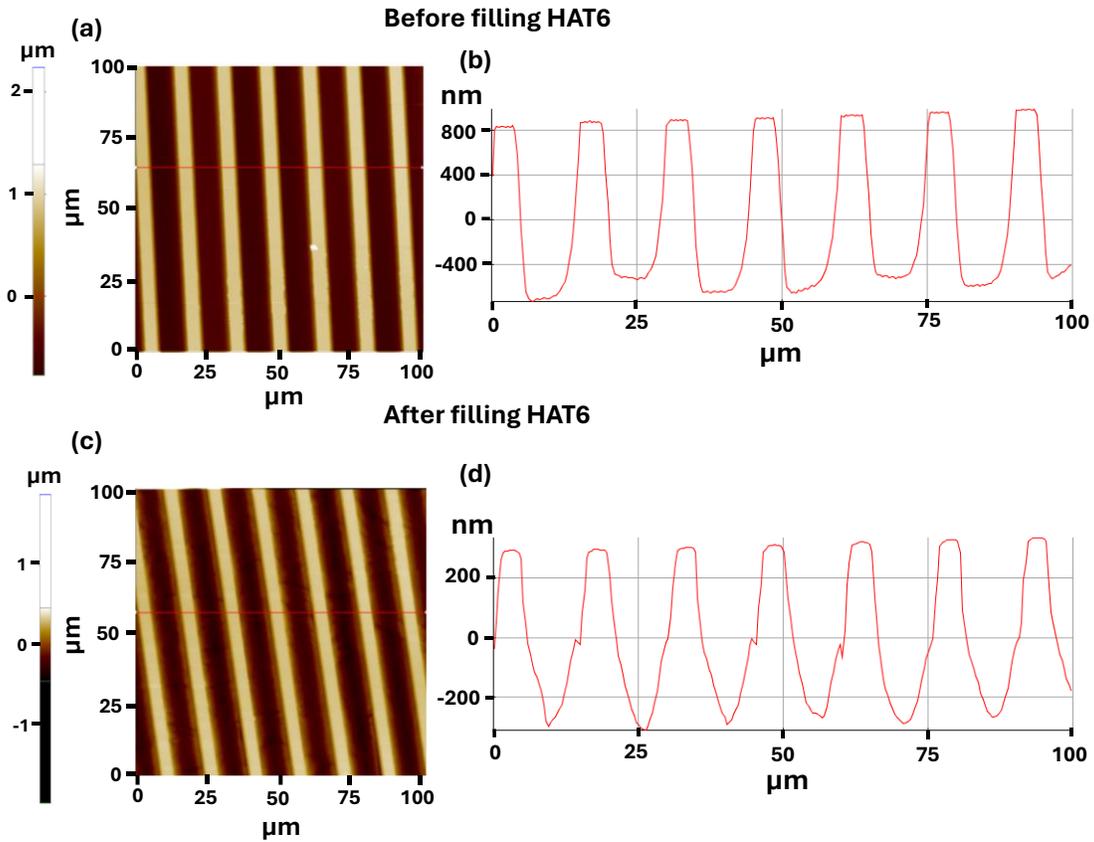

***Figure S1:*** *AFM images and linecuts of microchannel before filling (a, b) and after filling (c, d) of HAT6. The difference between the two heights determines the thickness of the film.*

**Section S2: Estimation of cooling-rate during quenching**

To estimate the cooling rate associated with transferring the sample from the Linkam hot stage to a copper block, the in-plane anisotropy parameter for known and controlled cooling rates is fit to a power-law function of the form $y = a + b \cdot x^c$. The cooling rates shown in **Figure S2** are accessed through a THMS600-H Linkam hot stage. The in-plane anisotropy parameter is measured for crystalline HAT6 prepared at different cooling rates using polarized optical microscopy as described in the main text. The fit parameters obtained from the curve are a = –1.67, b = 1.44, and c = 0.08. Using the experimentally measured anisotropy value of 0.7 for the quenched sample, we solve the power law equation with the above fit parameters to determine the cooling rate. Based on this method, we estimate that the cooling rate associated with rapid quenching is approximately 495 °C min$^{-1}$.

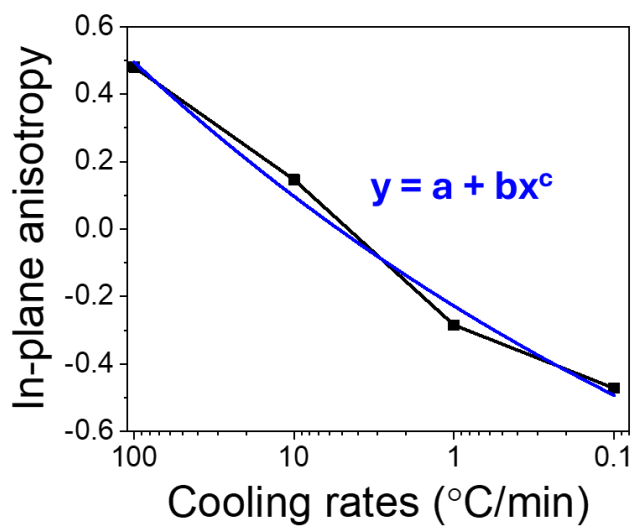

*Figure S2: Determination of the cooling rate associated with quenching on a copper block. The in-plane anisotropy parameter is plotted as a function of cooling rate. The data is fit to a power-law function. The fit equation and the experimentally measured anisotropy value are used to estimate the cooling rate associated with quenching.*

**Section S3: Calorimetric characterization of HAT6**

The phase sequence and phase transition temperatures of powdered HAT6 were determined using both conventional differential scanning calorimetry (DSC) and Flash DSC. Conventional DSC measurements were carried out using a TA Instruments DSC 250, with both heating and cooling scans performed at a rate of 10 °C/min. The phase sequence during heating is as follows: Cr 67.5 °C Col$_H$ 93.7 ° C I (Figure S3)

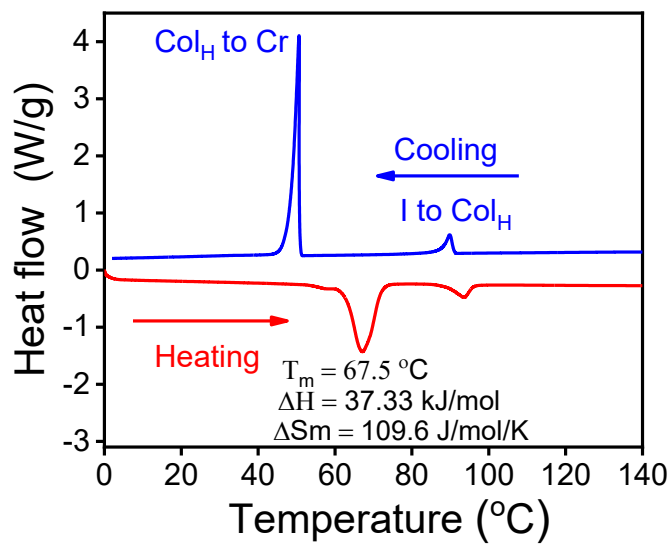

*Figure S3:* *Differential scanning calorimetry profile of HAT6 recorded during heating and cooling cycles at a rate of 10 °C/min. The data were fitted using TRIOS software, and the extracted thermal parameters - including the melting temperature ($T_m$), enthalpy of fusion ($\Delta H$), and entropy of fusion ($\Delta S$) are indicated in the plot.*

As only a limited range of cooling rates can be accessed during conventional DSC, we perform Flash DSC (Mettler Toledo) on a powder of HAT6 mounted on a MEMS chip. Flash DSC (see Figure S4) was performed over a wide range of cooling rates from 0.5 to 1,000 K/s. For the cooling cycle, the sample is first heated at 500 K/s and subsequently cooled at different cooling rates (Figure S4). The $Col_H$ → Crystal transitions shift to significantly lower temperatures with increasing cooling rate. This indicates that HAT6 can be supercooled at-least ≈ 50 K below the melting point under rapid cooling.

In microchannel experiments, the maximum supercooling we can attain is ≈ 30 K. On supercooling beyond this, as Flash DSC establishes is possible, we expect further reduction in the energy barrier separating the liquid crystalline and crystalline phases. On fast enough cooling, it might be possible to access a temperature range where the growth front velocity is temperature independent, and the transition is truly barrierless.

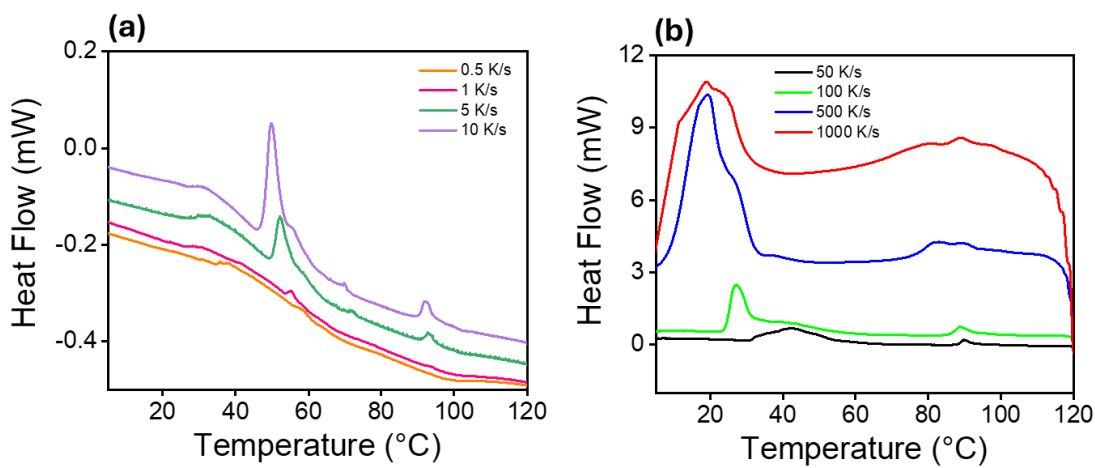

*Figure S4:* Flash DSC thermograms of HAT6 recorded during the cooling cycle. The sample was first heated at 500 K/s and subsequently cooled at various rates. For clarity in identifying the transition temperatures, the cooling rates are presented in two groups: (a) 0.5, 1, 5, and 10 K/s, and (b) 50, 100, 500, and 1000 K/s.

**Section S4: Reproducibility tests of crystalline alignment in samples produced from quenching of anchored Col$_H$ phase**

POM images of the crystalline state, acquired at 0° and 45° rotations of the channel relative to the crossed polarizers during quenching, are shown in **Figure S5** for three independent samples. All samples exhibit a clear change in intensity between the two orientations, confirming the retention of biaxiality in the crystalline state.

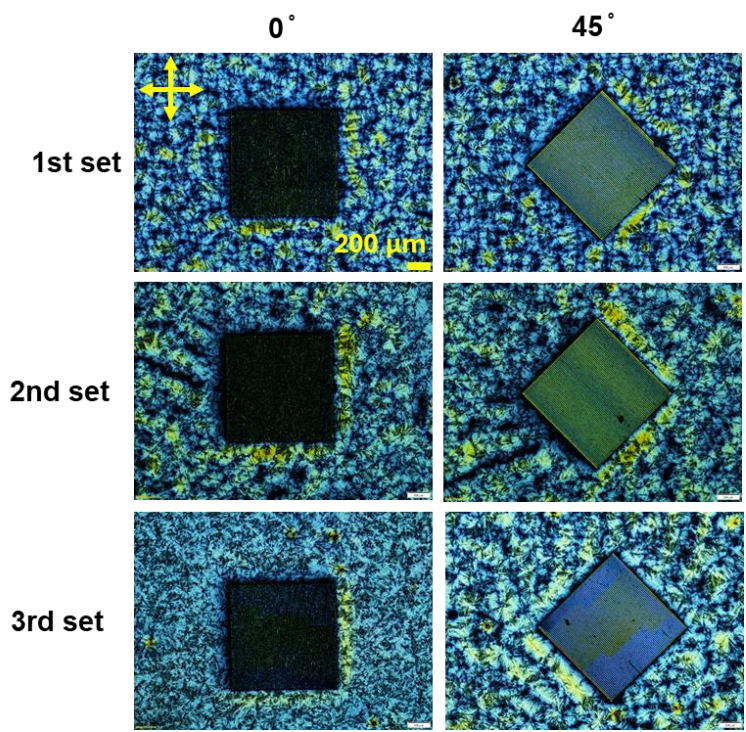

*Figure S5:* *Retention of biaxiality in the crystalline state. POM images captured at 0° and 45° orientations in the crystalline state during quenching from the Col$_H$ phase, for three independent sample sets with anisotropy values of 0.67, 0.67, and 0.71, respectively.*

**Section S5: POM images in the Col$_H$ and crystalline states of HAT6 confined in circular microchannels**

POM images of HAT6 confined within concentric microchannels in both the liquid crystalline Col$_H$ and crystalline states (during quenching) are shown in **Figure S6**. The alternating dark and bright regions (spherulitic morphology) observed in both phases indicate that the retention of orientation correlations between the precursor and transformed phases is independent of the channel geometry.

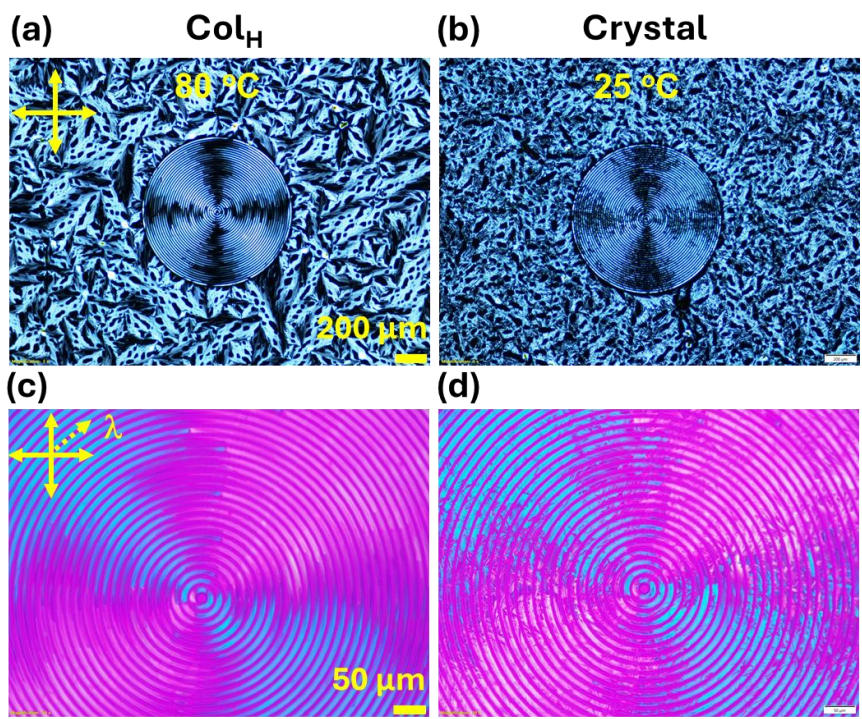

*Figure S6:* *POM images of HAT6 in circular microchannels in the liquid crystalline Col$_H$ (a) and crystal states (b). Panels (c) and (d) display the corresponding POM images taken with a compensator; the arrow indicates the orientation of the λ-plate relative to the crossed polarizers. The central circular region corresponds to the patterned area, while the surrounding region is the un-patterned background. The dark regions occur due to the alignment of the director along the cross-polarizers. The banded morphology created in the liquid crystal state is transferred to the crystal state through quenching.*

**Section S6: Order-preserving reversible phase transitions between liquid-crystalline Col$_H$ and crystalline states**

POM images (Figure S7) of the channels oriented along and at 45° to the polarizers in the Col$_H$ phase (80 °C) clearly show changes in birefringence, confirming in-plane alignment or biaxiality (red dotted line in Figure S7b). This biaxiality is retained in the crystalline state (25 °C) during rapid cooling at 100 °C/min (green dotted line in Figure S7b). Furthermore, re-heating from the crystalline to the Col$_H$ state also preserves the biaxiality (Figure S7a). Additionally, the degree of anisotropy is independent of the heating rate (second heating) as shown in (b). A slight reduction in overall anisotropy during the second heating may be attributed to tempering effects.

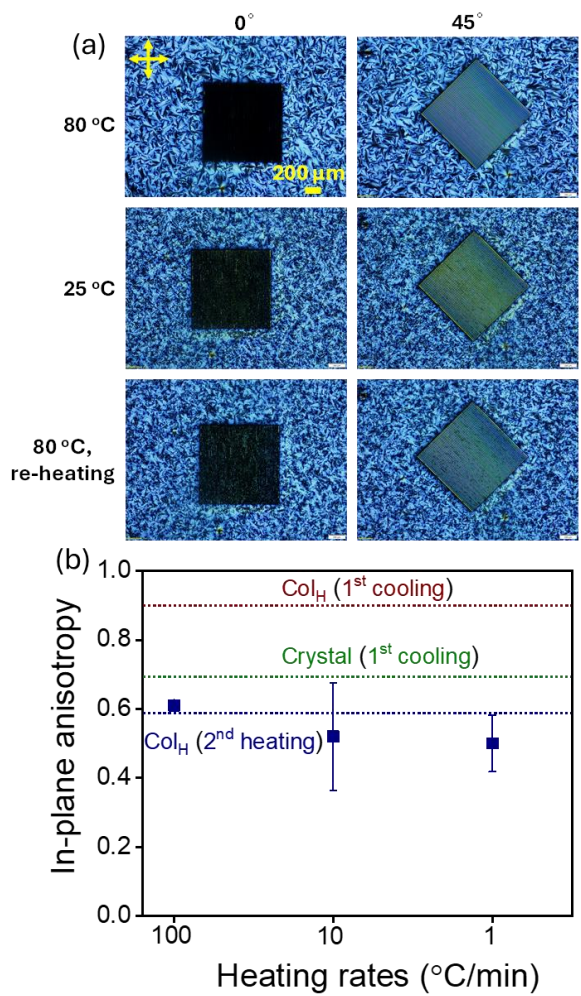

*Figure S7:* *Reversibility of Col$_H$-crystal transition in microchannels. Polarized optical microscopy images with the channels oriented along and at 45° with respect to the polarizers in the Col$_H$ phase on first cooling from the isotropic liquid, the crystal state on quenching the sample, and reheating the quenched sample (a). Panel (b) shows the in-plane anisotropy of the Col$_H$ phase on reheating the quenched crystal. The initial planar alignment of the Col$_H$ phase is largely retained. Values are averaged over three independent samples; error bars represent the standard deviation.*

**Section S7: Isothermal crystal growth kinetics at fastest supercooling temperature**

Crystal growth kinetics under isothermal conditions ($T_s = 40\ °C$, corresponding to the temperature of maximum supercooling) are shown for four independent microchannels. Sequential images were extracted at regular time intervals (20 frames/sec), and the crystal growth was quantified as a function of time. The plots exhibit predominantly linear increase in crystal size with time, interspersed with discrete jumps, rather than evolving smoothly. Such behavior is characteristic of systems undergoing intermittent, collective rearrangements and is commonly associated with phenomena such as seismic activity.

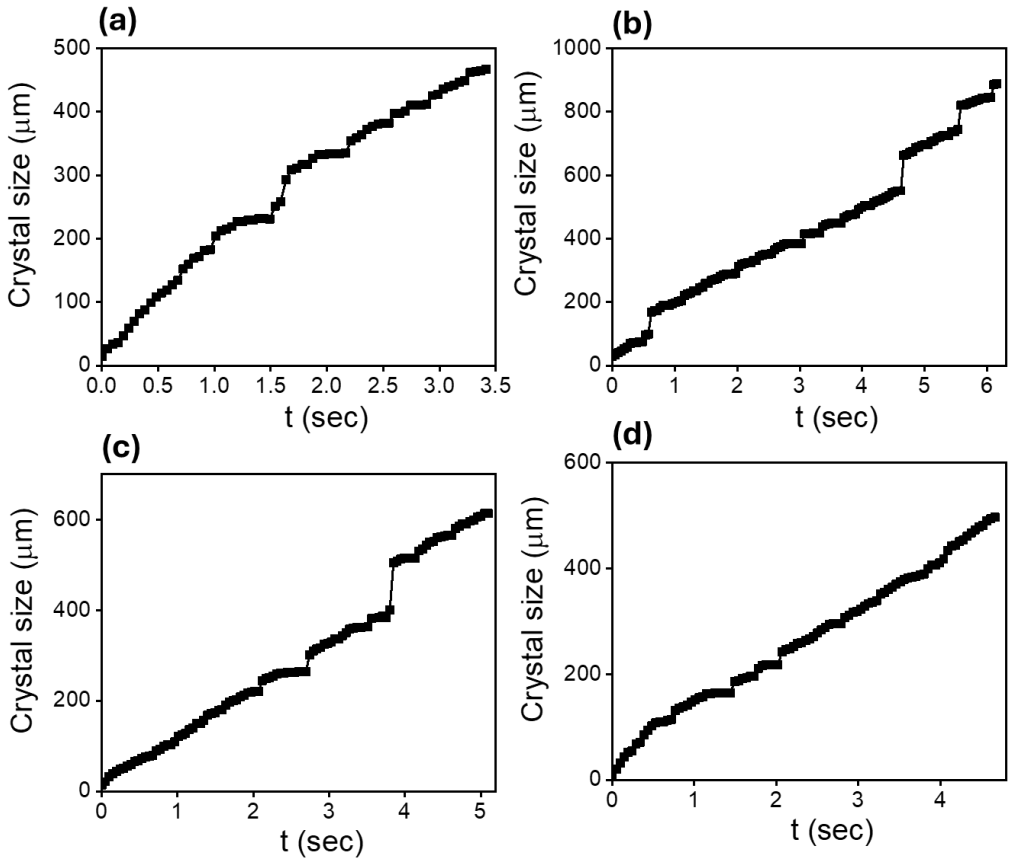

*Figure S8:* *Isothermal crystal growth kinetics during the Col$_H$ → Crystal transition under rapid supercooling conditions (40 °C). Crystal size as a function of time is shown for four independent channels (a-d). The growth profiles display a predominantly linear increase in crystal size with time, interspersed with discrete jumps characteristic of avalanche-like kinetics.*

**Section S8: Effect of cooling rate on transition temperature**

The Col$_H$-to-crystal phase transition is monitored under POM at cooling rates of 100, 10, 1, and 0.1 °C min$^{-1}$. A reduction in the transition temperature is observed at higher cooling rates, with a pronounced temperature drop due to rapid cooling at 100 °C min$^{-1}$ (Figure S9).

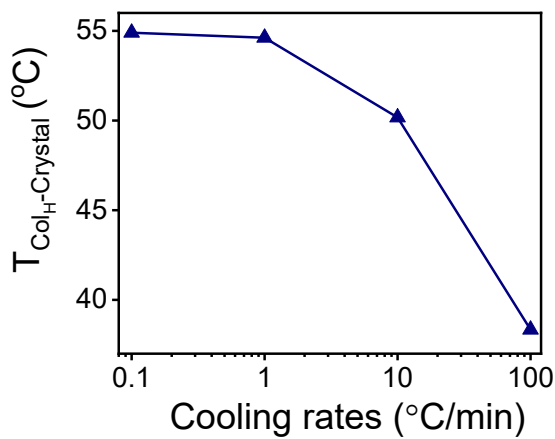

*Figure S9: Col$_H$→ Crystal (T$_{ColH-Crystal}$) transition temperatures as a function of cooling rates show a drop of transition temperature with higher cooling rates.*

**Section S9: Effect of degree of supercooling on order preservation**

To understand the effect of supercooling temperature on the order preservation in the crystalline state, POM images are captured with channels oriented along and at 45° to the polarizers and are shown in **Figure S10**. The sample is cooled from the aligned Col$_H$ phase (80 °C) rapidly at a rate of 100 °C/min and then held at different temperatures below the melting point (Crystal→ Col$_H$). The images acquired at different supercooling temperatures clearly show that in-plane anisotropy decreases as the supercooling temperature approaches the melting point and shows a negative value when the temperature is closer to the phase transition (Figure 4b). This observation indicates that large supercooling reduces the energy barrier, making a Martensitic-like transformation mechanism more favorable.

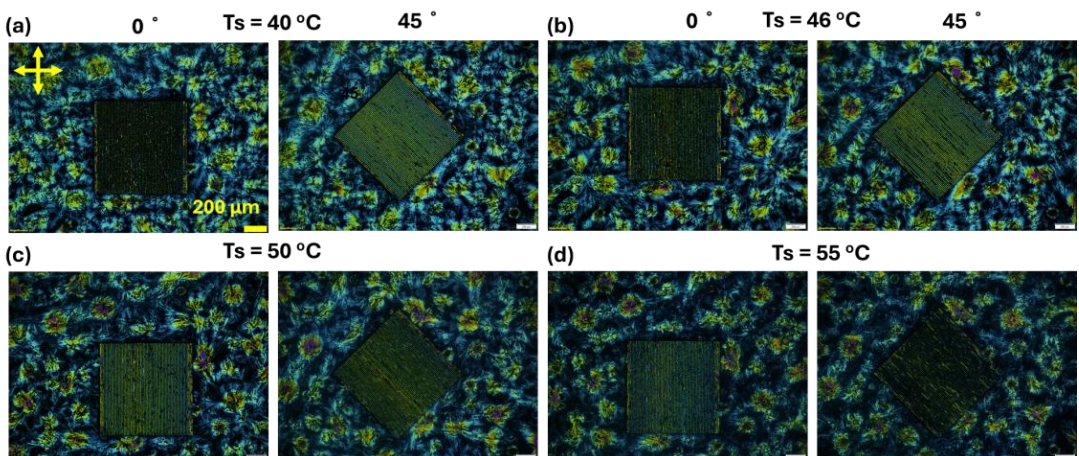

*Figure S10: Polarized optical microscopy images with the channels oriented 0° and at 45° with respect to the polarizers in the in the crystalline state held at different supercooling temperatures.*

**Section S10: GIWAXS patterns in the crystalline state obtained from slow cooling of aligned liquid crystal**

GIWAXS patterns of HAT6 films in the crystalline state, obtained by cooling from the $Col_H$ phase at rates of 10 and 0.1 °C min$^{-1}$, are shown in **Figure S11**. The patterns are collected in two measurement configurations, schematically illustrated in panels (a,b), with the X-ray beam oriented parallel (c, e) and perpendicular (d, f) to the channel direction. In contrast to the quenched sample (Figure 5), which exhibits strong π-π stacking features when the X-ray beam is parallel to the channel and pronounced intercolumnar features when the beam is perpendicular, films cooled at slower rates do not show these characteristics. This indicates that the alignment order present in the precursor liquid crystalline phase is lost during slow cooling.

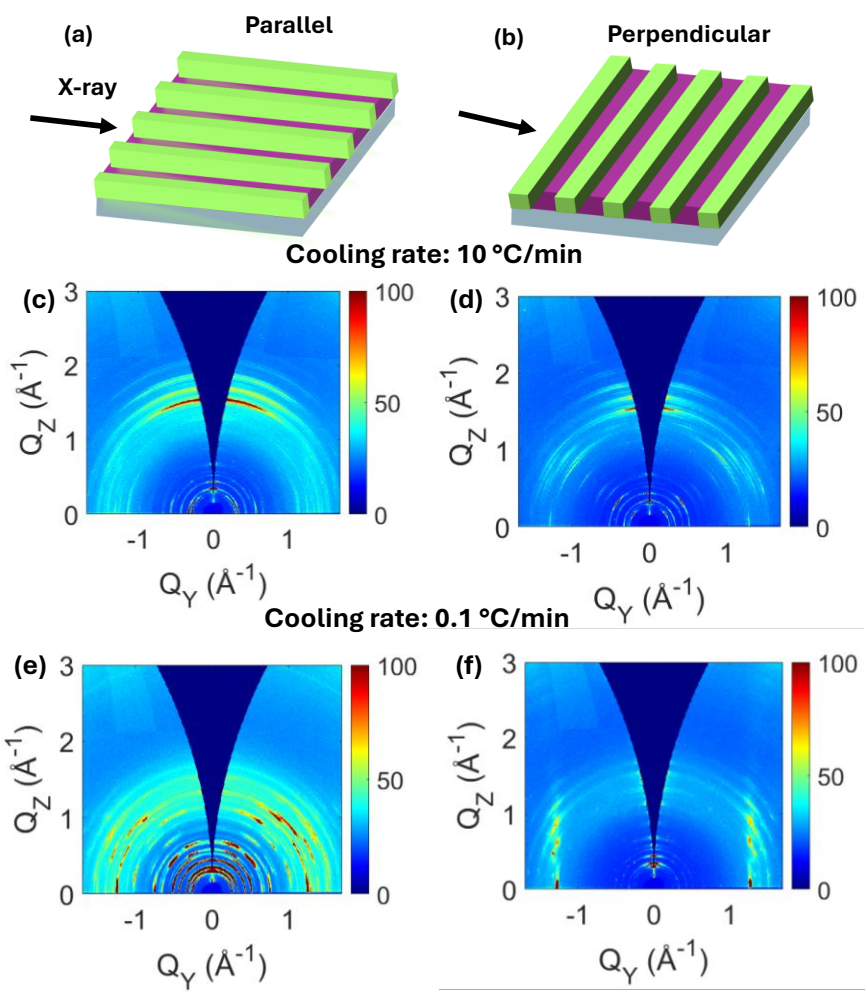

*Figure S11:* Grazing-incidence wide-angle X-ray scattering (GIWAXS) patterns in the crystalline state of HAT6-filled microchannel. Measurement configurations, schematically illustrated in (a, b). The patterns are acquired at cooling rates of 10 °C min$^{-1}$ (c, d) and 0.1 °C min$^{-1}$ (e, f). Measurements are performed with the microchannels oriented parallel (0°; c, e) and perpendicular (90°; d, f) to the incident X-ray beam.

**Section S11: Integration of 2D GIWAXS patterns**

Two-dimensional GIWAXS patterns in the crystalline state of quenched film are integrated over a Q range of 0 to 1.8 Å$^{-1}$ and a χ range of 70 to 80°, with the X-ray beam oriented parallel and perpendicular to the channel direction. The resulting intensity profile (I(Q) vs Q) is plotted and shown in **Figure S12**. A strong π-stacking feature is observed when the X-ray beam is oriented parallel to the channel wall, whereas this feature nearly disappears when the beam is perpendicular to the wall, indicating the retention of in-plane anisotropy in the crystalline state upon quenching from the aligned Col$_H$ phase.

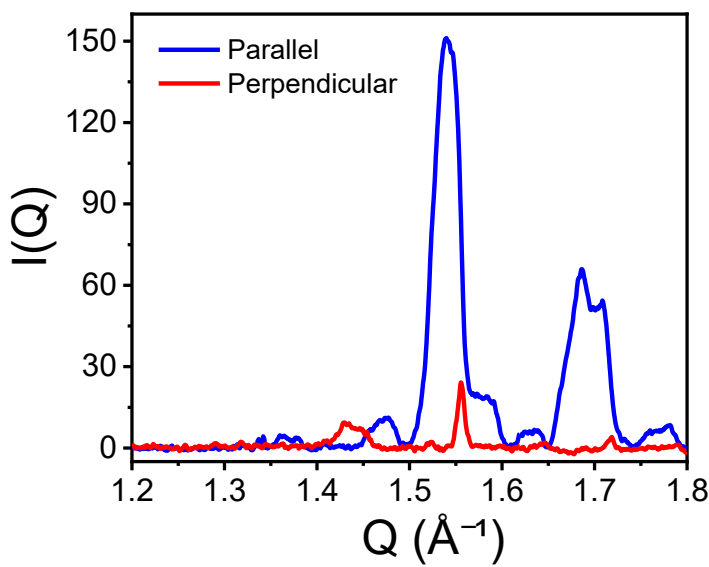

*Figure S12:* *Integrated 2D GIWAXS patterns of quenched HAT6 crystals. The corresponding I(Q) versus Q profile shows strong π-π stacking peaks when the X-ray beam is oriented parallel to the channel walls, whereas these peaks nearly disappear in the perpendicular configuration, confirming biaxial alignment in the crystalline state.*

**Section S12: Viscoelastic properties of liquid crystal and crystal states**

Strain sweep measurements were performed in both the liquid-crystalline Col$_H$ and crystalline states at a constant angular frequency of 5 rad s$^{-1}$ (Figure S13). In both states, the storage modulus (G′) exceeds the loss modulus (G″) within the linear viscoelastic regime, indicating predominantly elastic behavior. Beyond critical strain, G′ begins to decrease, marking the onset of structural yielding and the breakdown of the elastic network.

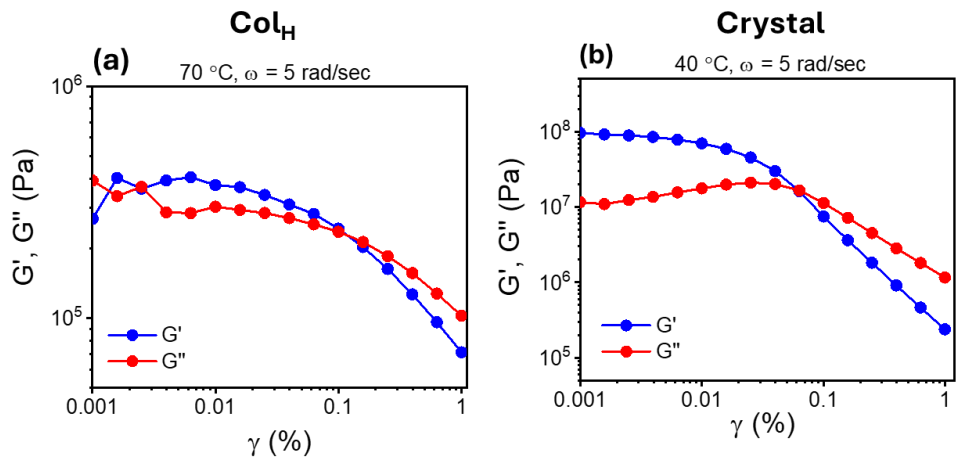

*Figure S13:* Storage and loss modulus as a function of oscillation strain of HAT6 in its liquid crystalline Col$_H$ (a) and crystalline states (b).

**Section S13: Crystallization from nematic and smectic liquid crystals**

LC test cells were fabricated using a clean silicon substrate (bottom) and a cover glass (top). A 10 wt% aqueous solution of polyvinyl alcohol (PVA) was spin-coated onto both substrates, followed by unidirectional rubbing to produce planar anchoring. The cells were assembled with the PVA-coated surfaces facing each other. Each cell had an active area of approximately 1 × 1 cm², and the cell thickness was defined by ~6 μm thick polyethylene terephthalate (PET) spacers. Both 5CB and 8CB samples were injected into the LC test cells by first heating them above their melting points, after which the isotropic melt filled the cells through capillary action.

Both 5CB and 8CB-filled LC cells were heated above their isotropic transition temperature (50°C) for 30 s to ensure complete melting. The samples were cooled from isotropic to liquid crystal at a controlled rate of 10 °C/min. In the liquid crystal, the sample was rotated in the plane to confirm the presence of in-plane ordering. Subsequently, the transition from the liquid-crystalline phase to the crystalline phase was induced by rapid cooling at 100 °C/min to determine whether any liquid-crystalline order is preserved in the crystalline state. The phase sequences for the materials are as follows:

5CB: Nematic (-22 °C) $\xrightarrow{100\ °C/min}$ Crystal (-30 °C)

8CB: Smectic A (0 °C) $\xrightarrow{100\ °C/min}$ Crystal (-10 °C)

To determine whether the transfer of biaxiality from the LC to the crystalline state is a general phenomenon or specific to certain LC orders, we studied crystallization from nematic and smectic phases. Both 5CB and 8CB were confined in PVA-coated cells, which provide planar surface anchoring for the LC molecules. As a result, a change in birefringence is observed with the director aligned along and 45° relative to the crossed polarizers. However, when the samples were rapidly cooled from both the nematic (Figure S14) and smectic (Figure S15) phases at 100 °C min$^{-1}$, no change in birefringence is observed between 0° and 45°. These results indicate that the LC order (nematic, with no translational order and smectic, with translational order in one direction) is not retained in the crystalline state, in contrast to discotic LCs, which exhibit translational order in two dimensions and preserve orientational correlations in the crystalline state.

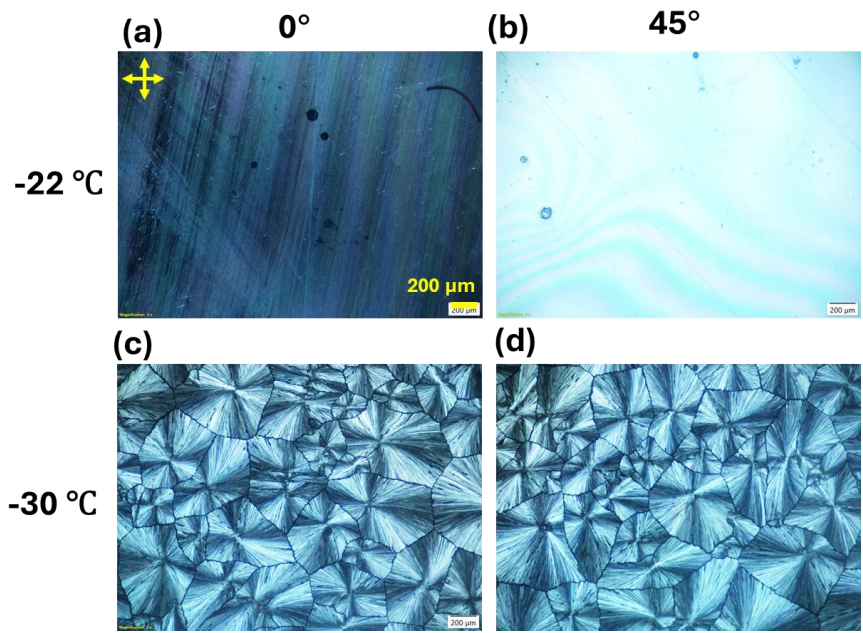

***Figure S14:*** *Crystallization from the nematic phase. POM images recorded at 0° and 45° in the nematic phase (a, b), and in the crystalline state obtained from the aligned nematic phase during rapid cooling at 100 °C/min (c,d). A clear change in birefringence is observed between 0° and 45° in the liquid-crystalline state, whereas no change in birefringence is observed in the crystalline state.*

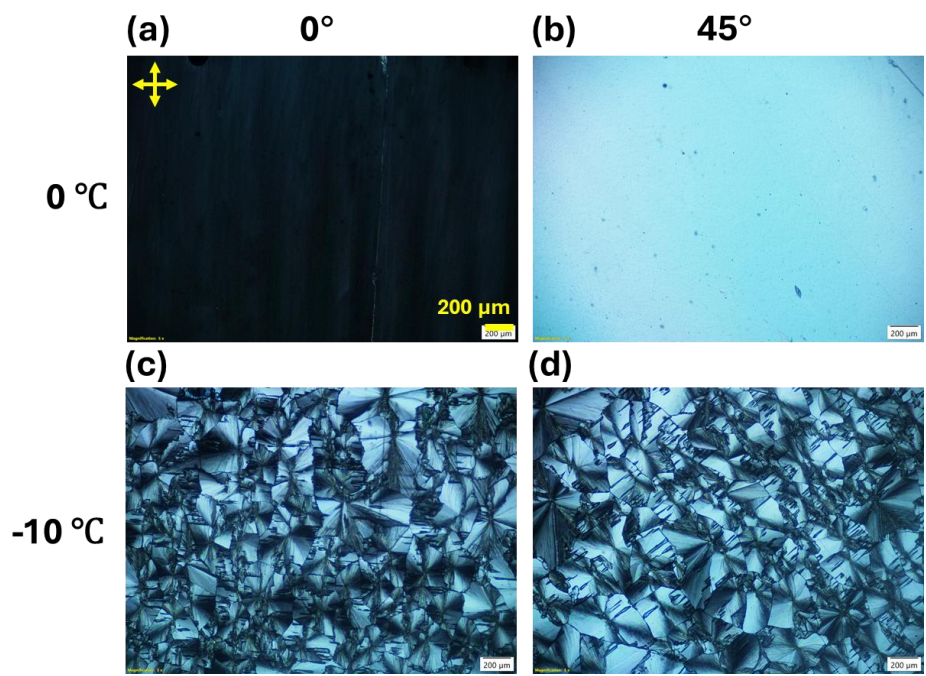

*Figure S15:* Crystallization from the smectic phase. POM images recorded at 0° and 45° in the smectic phase (a, b), and in the crystalline state obtained from the aligned smectic phase during rapid cooling at 100 °C/min (c, d). A clear change in birefringence is observed between 0° and 45° in the liquid-crystalline state, whereas no change in birefringence is observed in the crystalline state.